\documentclass [12pt]{iopart}
\usepackage{iopams}
\usepackage{color,graphicx}

\begin{document}
\title[Threshold of parametric decay instability\dots]{Threshold of parametric decay instability accompanying electron
Bernstein wave heating in spherical tokamaks}
\author{E Z Gusakov and A V Surkov}
\address{Ioffe Institute, Politekhnicheskaya 26,
194021 St. Petersburg, Russia} \eads{
\mailto{a.surkov@mail.ioffe.ru}}
\begin{abstract}
 The parametric instability of upper hybrid wave decay into back scattered
  upper hybrid wave and lower hybrid wave is considered for conditions of inhomogeneous plasma
  of spherical tokamaks. The possibility of absolute instability is demonstrated and the corresponding threshold
  is determined. It is shown that the threshold power increases
with pump frequency and electron temperature. Threshold power is
estimated for typical parameters of experiment in MAST tokamak. It
is shown that in this case parametrical reflection arises, if
probing power exceeds $90~\mbox{W}/\mbox{cm}^2$, which gives $30$~kW
for a beam of 10 cm radius.
\end{abstract}
%
\section{Introduction}
In recent years, considerable attention of the controlled fusion
community has been paid to spherical tokamaks (ST). These are
small aspect ratio devices with typically high plasma density and
comparatively low magnetic field. This ST feature has strong
effect on the electromagnetic wave propagation. In the microwave
frequency region, characteristic surfaces, like the upper hybrid
resonance and the cut-off are very close to the plasma edge. As a
result, the electromagnetic (EM) waves are unable to penetrate
into the plasma interior. The only way to overcome this difficulty
is to use the linear conversion of the incident EM wave into the
electron Bernstein wave (EBW) occurring in the upper hybrid
resonance (UHR). The latter has no density limitations and can, in
principle, carry the radio frequency power deep into the plasma.
This mechanism of wave conversion has been successfully
demonstrated to produce heating in over dense plasmas in the W7-AS
stellarator~\cite{W7AS}. The plasma heating experiment based on
this scheme is in progress now in the MAST tokamak at Culham, UK.
The wave propagation, conversion in the UHR and absorption is
usually accompanied, in 100~kW power level experiments, by
nonlinear effects, in particular, by parametric decay
instabilities (Versator~\cite{Versator}, \mbox{FT-1}~\cite{FT1},
W7-AS~\cite{W7AS}). These instabilities lead to redistribution of
incident power between plasma species and can cause anomalous
reflection, especially when excited at the plasma edge.

The present paper is devoted to analysis of the  decay instability
thresholds and growth rates for specific conditions of low
magnetic field typical for ST. The study is focused upon the decay
of UH wave into another UH wave and intermediate frequency range
wave satisfying the lower hybrid resonance condition, which was
observed in the UHR heating experiments mentioned above. The
influence of plasma inhomogeneity on its threshold is investigated
for  backscattering of the incident UH wave. Dependence of the
decay instability threshold on the pump frequency, necessary for
the heating experiment optimization, is studied.

The paper is organized as follows. In~\sref{sec:eqs} we deduce
equations,  describing the decay of the incident high-frequency UH
wave into UH wave and low-frequency LH wave:
$\ell_{UH}\to\ell_{UH}'+\ell_{LH}$, and consider them in WKB
approximation. In~\sref{sec:threshold} we calculate an absolute
instability threshold, which corresponds to UH wave induced back
scattering instability. Brief discussion follows
in~\sref{sec:disc}.
\section{Equations for wave amplitudes}
\label{sec:eqs} We use slab plasma model, i.e. density and
magnetic field gradients are assumed to be along $x$ axe. A
magnetic field direction is chosen along $z$ axe.
 We consider one dimensional problem of pump wave parametric decay.
 UH pump wave is supposed to be excited by external antenna via
tunnelling effect (X$\to$B scheme according to~\cite{Lashmore}),
and assumed to propagate in $x$ direction. We consider here the
high pump frequency case, when the frequency is larger than
doubled electron cyclotron frequency, corresponding to the
magnetic field in UHR: $\omega_0>2\omega_{ce}$. In this case the
UH pump wave  dispersion curve (see~\fref{fig:dispers}) does not
possess a turning point and transformation to Bernstein wave
occurs without change of group velocity sign.

By indices $_{0}$, $_{1}$, $_{2}$ we will mark  frequency,
wavenumber, complex amplitudes of the electric fields and
potentials of the pump wave, parametrically reflected UH wave and
LH wave correspondingly.
\subsection{Nonlinear current and equation for LH wave}
Poisson equation for LH waves can be represented in the following
form~\cite{Golant}
\begin{equation}
{\rm div} \vec D_{LH}=\frac{\rmd}{\rmd
x}\left[\left(\varepsilon(\omega_2)+\ell_T^2(\omega_2)\frac{\rmd^2}{\rmd
x^2}\right)E^{LH}_x\right]+\eta(\omega_2)\frac{\rmd E^{LH}_z}{\rmd
z}=4\pi\rho_{LH}\label{eq:Poisson}
\end{equation}
Here $\vec E^{LH}=-\nabla\phi_{LH}$ is an electric field of LH
wave, which is assumed to be potential, $\varepsilon$, $\eta$ are
the components of the dielectric tensor
\[
\varepsilon(\omega)=1-\frac{\omega_{pe}^2}{\omega^2-\omega_{ce}^2}-\frac{\omega_{pi}^2}{\omega^2-\omega_{ci}^2},\qquad
\eta(\omega)=1-\frac{\omega_{pe}^2}{\omega^2}
\]
which for LH wave frequency
$\omega_2\sim\sqrt{\omega_{ce}\omega_{ci}}$ take the form
\[
\varepsilon(\omega_2)\simeq1+\frac{\omega_{pe}^2}{\omega_{ce}^2}-\frac{\omega_{pi}^2}{\omega_2^2},\qquad
\eta(\omega_2)\simeq-\frac{\omega_{pe}^2}{\omega_2^2}
\]
Parameter $\ell_T$ is associated with particles thermal
motion~\cite{Akhiezer}
\begin{equation}
\ell_T^2(\omega)=\frac32\left(\frac{\omega_{pe}^2}{\omega^2-\omega_{ce}^2}\frac{V_{Te}^2}{\omega^2-4\omega_{ce}^2}+
\frac{\omega_{pi}^2}{\omega^2-\omega_{ci}^2}\frac{V_{Ti}^2}{\omega^2-4\omega_{ci}^2}\right)\label{eq:LT2}
\end{equation}
 where $V_{Te,i}$
corresponds to electron and ion thermal velocity
$V_{Te,i}=\left(2T_{e,i}/m_{e,i}\right)^{1/2}$. In particular, for
LH wave~\eref{eq:LT2} takes the form
\[
\ell_T^2(\omega_2)\simeq\frac32\frac{\omega_{pe}^2}{\omega_{ce}^4}\left(\frac14V_{Te}^2+\frac{m_i}{m_e}V_{Ti}^2\right)
\]

 Thus, equation~(\ref{eq:Poisson}), describing the excitation
of LH wave, can be rewritten as
\begin{equation}
{\rm div} \vec
D_{LH}=-\ell_T^2(\omega_2)\phi_{LH}''''-\varepsilon(\omega_2)\phi''_{LH}-\varepsilon'(\omega_2)\phi'_{LH}+k_z^2\eta(\omega_2)\phi_{LH}=4\pi\rho_{LH}
\label{eq:Poisson1}
\end{equation}
Here and below $\phantom{\phi}'$ denotes $\rmd/\rmd x$. A charge
density $\rho$ is associated with nonlinear current $j_{LH}$ by
continuity equation
\[
\frac{\partial\rho_{LH}}{\partial t}+\frac{\partial
j_{LH}}{\partial x}=0
\]
To obtain nonlinear current $j_{LH}$ we consider electron motion
in the field of three potential waves
\begin{equation}
\left\{ \eqalign{
\dot v_x&=-\frac{e}{2m_e}\sum_{i=0,1,2}\left\{E_i\exp\left[\rmi \int^x k_i(x')\rmd x'-\rmi\omega_it\right]+{\rm c.c.}\right\}-\omega_{ce}v_y\\
\dot v_y &=\omega_{ce}v_x\\
} \right. \label{eq:motion}
\end{equation}
Here dot $\dot{\phantom{v}}$ means $\rmd/\rmd t$. Electric field
of three waves is taken in geometrical optics (or WKB)
approximation. This approximation is not valid if the decay
point~$x_d$, determining in the inhomogeneous plasma by the
conditions
\begin{equation}
k_0(x_d)=k_1(x_d)+k_2(x_d),\qquad \omega_0=\omega_1+\omega_2
\label{eq:Bragg}
\end{equation}
 is
situated in the vicinity of LH wave turning point
(see~\sref{sec:disc} for proper discussion of corresponding
criteria).

In deducing~\eref{eq:Poisson} we assumed following criteria to be
satisfied
\[
\frac{k_2^2V_{Te}^2}{\omega_{ce}^2}\ll1,\quad
\frac{k_z^2V_{Te}^2}{\omega_2^2}\ll1,\quad\left|\frac{\omega_2-\omega_{ce}}{k_zV_{Te}}\right|\gg1,
\quad\left|\frac{\omega_2-2\omega_{ce}}{k_zV_{Te}}\right|\gg1
\]
First criterion, which characterizes $k\rho$-approximation, allows
us to get nonlinear component of a solution of~\eref{eq:motion} in
the form
\begin{eqnarray*}
\fl
v_{LH}=\frac{e^2}{4m_e^2}\sum_{{i,j=0,1,2}}\frac{k_i}{\omega_j^2-\omega_{ce}^2}\left\{
\frac{\omega_i+\omega_j}{(\omega_i+\omega_j)^2-\omega_{ce}^2}E_iE_j\exp\left[\rmi\int^x
(k_i+k_j)\rmd
x'-\rmi(\omega_i+\omega_j)t\right]\right.\\\lo{+}\left.\frac{\omega_i-\omega_j}{(\omega_i-\omega_j)^2-\omega_{ce}^2}E_iE_j^*
\exp\left[\rmi\int^x (k_i-k_j)\rmd
x'-\rmi(\omega_i-\omega_j)t\right]+{\rm c.c.}\right\}
\end{eqnarray*}
Averaging $v_{LH}$ we neglect high-frequency terms. That yields
\begin{eqnarray*}
\fl \left\langle v_{LH}\right\rangle=\frac{e^2}{4m_e^2}
\frac{\omega_1-\omega_0}{(\omega_0-\omega_s)^2-\omega_{ce}^2}
\left(\frac{k_0}{\omega_1^2-\omega_{ce}^2}-\frac{k_1}{\omega_0^2-\omega_{ce}^2}\right)
E_0E_1^*\\\lo{\times}\exp\left[\rmi\int^x (k_0-k_1)\rmd
x'-\rmi(\omega_0-\omega_1)t\right] +{\rm c.c.}
\end{eqnarray*}
Taking into account that $j_{LH}=-en\left\langle
v_{LH}\right\rangle$ and passing to the complex amplitudes of the
potential $\phi_{LH}=(\phi_2+\phi_2^*)/2$ one obtains an equation
for LH wave
\begin{eqnarray}
\fl-\ell_T^2(\omega_2)\phi_2''''-\varepsilon(\omega_2)\phi''_2-\varepsilon'(\omega_2)\phi'_2+k_z^2\eta(\omega_2)\phi_2
\nonumber\\\lo{=} \frac{e}{2m_e}
\frac{\omega_{pe}^2}{\omega_{ce}^2}k_0k_1(k_0-k_1)
\left(\frac{k_0}{\omega_1^2-\omega_{ce}^2}-\frac{k_1}{\omega_0^2-\omega_{ce}^2}\right)
\phi_0\phi_1^*\nonumber\\\times\exp\left[\rmi\int^x (k_0-k_1)\rmd
x'-\rmi(\omega_0-\omega_1)t\right]\label{eq:LH}
\end{eqnarray}

\subsection{Nonlinear current and equation for UH wave}
For UH waves we have~\cite{Golant}
\begin{equation}
\left\{\eqalign{&{\rm }D_{UH}=\frac{\rmd}{\rmd
x}\left[\left(\varepsilon(\omega_0)+\ell_T^2(\omega_0)\frac{\rmd^2}{\rmd
x^2}\right)E^{UH}_x+\rmi
g(\omega_0)E^{UH}_y\right]=4\pi\rho_{UH}\\
&\frac{\rmd^2}{\rmd x^2}E^{UH}_y=\rmi
g(\omega_0)\frac{\omega_0^2}{c^2}E^{UH}_x }\right.\label{eq:sysUH}
\end{equation}
Here corresponding components of dielectric tensor take the
following form for the frequency of UH wave
$\omega_1^2\approx\omega_{UH}^2=\omega_{pe}^2+\omega_{ce}^2$
\[
\varepsilon(\omega_1)=1-\frac{\omega_{pe}^2}{\omega_1^2-\omega_{ce}^2},
\qquad g(\omega_1)\simeq\frac{\omega_{ce}}{\omega_1}
\]
Parameter $\ell_T(\omega_1)$ can be represented as
\[
\ell_T^2(\omega_1)=\frac{3V_{Te}^2}{2(\omega_1^2-4\omega_{ce}^2)}<\ell_T^2(\omega_2)
\]
Considering potential UH wave $\vec E^{UH}=-\nabla\phi_{UH}$, and
substituting integrated second equation of~\eref{eq:sysUH} to the
first, one obtains
\begin{equation}
-\ell_T^2(\omega_1)\phi''''_{UH}-\varepsilon(\omega_1)\phi''_{UH}-
\varepsilon'(\omega_1)\phi'_{UH}+g^2(\omega_1)\frac{\omega_1^2}{c^2}\phi_{UH}=4\pi\rho_{UH}
\label{eq:Poisson2}
\end{equation}

A charge density $\rho_{UH}$ is associated with nonlinear current
$j_{UH}$ as
\[
\frac{\partial\rho_{UH}}{\partial t}+\frac{\partial \left\langle
j_{UH}\right\rangle}{\partial x}=0,\qquad j_{UH}=-e\delta
n_{\omega_2}v_{\omega_0}
\]
Here $v_{\omega_0}$ describes electron motion in the field of the
pump wave
\[
v_{\omega_0}=-\frac{1}{8\pi
ne}\frac{\omega_{pe}^2\omega_0}{\omega_0^2-\omega_{ce}^2}\left\{\rmi
E_0\exp\left[\rmi\int^x k_0(x')\rmd x'-\rmi\omega_0t\right]+{\rm
c.c.}\right\}
\]
Density modulation $\delta n_{\omega_2}$  is caused by the
electron motion in the field of LH wave
\begin{eqnarray*}
\delta n_{\omega_2}=-\frac1{8\pi
e}\frac{\omega_{pe}^2k_2}{\omega_2^2-\omega_{ce}^2}\left\{\rmi
E_2\exp\left[\rmi\int^x k_2(x')\rmd x'-\rmi\omega_2 t\right]+{\rm
c.c.}\right\}
\end{eqnarray*}
Here we omitted a contribution of LH wave component along the
magnetic field, which is smaller in factor of
$k_z^2V_{Te}^2/\omega_2^2\ll1$.

Averaging the nonlinear current, we leave the terms varying with
frequency $\omega_0-\omega_2$ only. That gives
\[
\fl\left\langle j_{UH}\right\rangle= \frac{1}{16\pi
}\frac{e}{m_e}\frac{\omega_{pe}^2}{\omega_{ce}^2}
\frac{\omega_0k_2}{\omega_0^2-\omega_{ce}^2}\left\{
E_0E_2^*\exp\left[\rmi\int^x (k_0-k_2)\rmd
x'-\rmi(\omega_0-\omega_2)t\right]+{\rm c.c.}\right\}
\]
yielding~\eref{eq:Poisson2} for the complex amplitude of the
potential $\phi_{UH}=(\phi_1+\phi_1^*)/2$ in the form
\begin{eqnarray}
\fl-\ell_T^2(\omega_1)\phi_1''''-\varepsilon(\omega_1)\phi''_1-\varepsilon'(\omega_1)\phi'_1
+g^2(\omega_1)\frac{\omega^2}{c^2}\phi_1\nonumber\\\lo{=}
\frac{e}{2m_e} \frac{\omega_{pe}^2}{\omega_{ce}^2}
\frac{k_0k_2^2(k_0-k_2)}{\omega_1^2-\omega_{ce}^2}
\phi_0\phi_2^*\exp\left[\rmi\int^x (k_0-k_2)\rmd
x'-\rmi(\omega_0-\omega_2)t\right]\label{eq:UH}
\end{eqnarray}
\subsection{WKB-analysis of the equations obtained}
\begin{figure}
\begin{center}
\includegraphics[height=0.2\textheight]{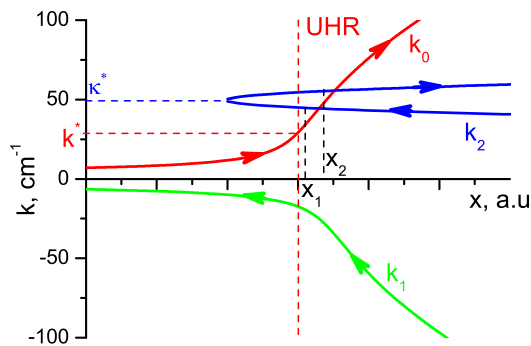}
\end{center}
\caption{\label{fig:dispers} UH and LH waves dispersion curves in
high frequency case ($\omega_{0}>2\omega_{ce}$).}
\end{figure}

A dispersion relations, which can be obtained from
equations~(\ref{eq:Poisson1}),~(\ref{eq:Poisson2}), when
$\rho_{UH}=\rho_{LH}=0$, take the following form~\cite{Golant}:
for UH waves
\begin{equation}
\varepsilon(\omega_{0,1})=\ell_T^2(\omega_{0,1})\left(k_{0,1}^2-\frac{k_*^4}{k_{0,1}^2}\right)
\label{eq:dispUH}
\end{equation}
where the transformation wavenumber is
\[
k_*^2=\frac{\omega_0}{c}\frac{g}{\ell_T(\omega_0)}=\frac{\omega_{ce}/c}{\ell_T(\omega_0)}
\]
and for LH waves
\begin{equation}
\varepsilon(\omega_2)=\ell_T^2(\omega_2)\left(k_2^2+\frac{\varkappa_*^4}{k_2^2}\right),\qquad
\varkappa_*^4=-\frac{\eta k_z^2}{\ell_T^2(\omega_2)}
\label{eq:dispLH}
\end{equation}
Equations~(\ref{eq:dispUH}),~(\ref{eq:dispLH}) allow us to obtain
group velocities of the corresponding waves. We get
\begin{equation}
\fl
v_{0,1}=\ell_T^2(\omega_{0,1})\frac{\omega_{0,1}^2-\omega_{ce}^2}{\omega_{0,1}}\left(k_{0,1}+\frac{k_*^4}{k_{0,1}^3}\right),\qquad
v_2=\ell_T^2(\omega_2)\frac{\omega_2\omega_{ce}^2}{\omega_{pe}^2}\left(k_2-\frac{\varkappa_*^4}{k_2^3}\right)
\label{eq:group}
\end{equation}
One can see
from~(\ref{eq:dispUH}),(\ref{eq:dispLH}),~(\ref{eq:group}) that in
the probing frequency range under consideration
$\omega_0>2\omega_{ce}$, which is used at present for EBW heating
in MAST, there is no  change of the group velocity sign in the UHR
point. The transformation point of LH wave, which is shifted from
LH resonance position (where $\varepsilon(\omega_2)=0$), is the
turning point of LH wave, and group velocity changes the sign
there. Corresponding dispersion curves are represented
in~\fref{fig:dispers}. We consider the most interesting case of
$k_0>0$, $k_1<0$, when the group velocity directions  give rise to
positive feedback loop, which can lead to absolute decay
instability~\cite{Rosenbluth,White,Piliya73,Piliya74}.

We consider equations~(\ref{eq:LH}),~(\ref{eq:UH}) in WKB
approximation, substituting
\[
\phi_{0,1,2}=\frac{a_{0,1,2}}{k_{0,1,2}\sqrt{v_{0,1,2}}}\exp\left[\rmi\int^x
k_{0,1,2}(x')\rmd x'-\rmi\omega_{0,1,2}t\right]
\]
and neglecting corresponding small terms. In the vicinity of the
decay point~(\ref{eq:Bragg}) we have
\begin{eqnarray}
\left\{ \eqalign{ {a_1^*}'&=\nu_1a_2\exp\left[-\rmi\int^x
(k_0-k_1-k_2)\rmd
x'\right]\\
a_2'&= \nu_2a_1^*\exp\left[\rmi\int^x
(k_0-k_1-k_2)\rmd x'\right] } \right. \nonumber\\
\nu_1=-\frac{e}{4m_e}\frac{\omega_{pe}^2}{\omega_1\omega_{ce}^2}
\frac{k_2a_0^*}{\sqrt{-v_0v_1v_2}},\qquad
\nu_2=\frac{e}{4m_e}\frac{\omega_2}{\omega_0^2-\omega_{ce}^2}
\frac{k_2a_0}{\sqrt{-v_0v_1v_2}}\label{eq:nu}
\end{eqnarray}

\section{Absolute instability threshold}
\label{sec:threshold} Absolute instability can arise, when decay
conditions~(\ref{eq:Bragg}) allow two decay points $x_{1,2}$ to
exist, and the group velocities directions provide positive
feedback loop. In this case, according
to~\cite{Piliya73,Piliya74}, the absolute instability threshold is
determined by the following conditions on the waves amplification
coefficient
\begin{equation}
\left|S_{12}(x_1)\right|\left|S_{21}(x_2)\right|=1\label{eq:S}
\end{equation}
where $S_{jk}(x_i)$ is the wave amplitude $a_k$, which leaves the
vicinity of decay point $x_i$, due to incidence onto this point of
the wave $a_j$ of unit amplitude
\begin{eqnarray*}
S_{12}=-\frac{\nu_2\ell\sqrt{-2\pi\rmi}}{\Gamma\left(\rmi|Z|+1\right)}\rme^{\pi|Z|/2},\qquad
S_{21}=\frac{\nu_1\ell\sqrt{2\pi\rmi}}{\Gamma\left(-\rmi|Z|+1\right)}\rme^{\pi|Z|/2}
\end{eqnarray*}
where $Z=\ell^2\nu_1\nu_2$ and $\ell$ is the length of the decay
region
\[
\ell^{2}(x_d)=\left.\left\{\frac{\rmd}{\rmd
x}\left[k_0(x)-k_1(x)-k_2(x)\right]\right\}^{-1}\right|_{x=x_d}
\]

 The spectrum of the instabilities arising is determined by
the condition on the phase, gained in the feedback loop
\begin{equation}
\Phi=\int_{x_1}^{x_2} \left[k_0(x')-k_1(x')-k_2(x')\right]\rmd x'
+\frac\pi2=2\pi N,\qquad N=1,2,\dots\label{eq:Phi}
\end{equation}

To calculate the decay instability threshold we act in accordance
with following procedure. We calculate the terms, involved
in~\eref{eq:S} using~\eref{eq:Phi}, and then substitute them
to~\eref{eq:S}, obtaining an equation for threshold power.

 At first we calculate the decay point coordinates $x_{1,2}$. It
should be noted that the UHR position $x_{UH}(\omega_1)$ of the
parametrically reflected UH wave is shifted  in respect to the UHR
position of the pump wave $x_{UH}(\omega_0)$. This shift can be
estimated as
\begin{eqnarray}
x_{UH}(\omega_1)-x_{UH}(\omega_0)=
\frac{2L\left[x_{UH}(\omega_0)\right]\omega_0\omega_2}{\omega
_{pe}^2\left[x_{UH}(\omega_0)\right]}\nonumber\\
L(x)=\left[\frac1n\frac{\rmd n(x)}{\rmd x}
 +\frac{2\omega_{ce}^2}{\omega_{pe}^2B}\frac{\rmd B(x)}{\rmd
 x}\right]^{-1}\label{eq:L}
\end{eqnarray}
where $n(x)$, $B(x)$ are plasma density and magnetic field
correspondingly. The decay points $x_1$,~$x_2$ are situated in the
vicinity of pump wave UHR resonance $x_{UH}(\omega_0)$.
 It can be shown that for real plasma parameters
the distance between UH resonances
$x_{UH}(\omega_1)-x_{UH}(\omega_0)$ is substantial to provide
\[
\left|k_1\left[x_{UH}(\omega_0)\right]\right|\sim\frac{\ell_T(\omega_0)k_*^2\omega_{pe}}{\sqrt{2\omega_0\omega_2}}\ll
k_*= \left|k_0\left[x_{UH}(\omega_0)\right]\right|
\]
This allows us to neglect $k_1(x_{1,2})$ in the decay
condition~(\ref{eq:Bragg}), writing it as
\begin{equation}
k_0(x_{1,2})=k_2(x_{1,2})\label{eq:eq}
\end{equation}
To solve this equation we assume the dielectric permeability
$\varepsilon$ to vary linearly in the region considered
\begin{eqnarray*}
\varepsilon(x,\omega_0)=\frac{x-x_{UH}(\omega_0)}{L\left[x_{UH}(\omega_0)\right]},\qquad
\varepsilon(x,\omega_2)=\frac{x-x_{LH}(\omega_2)}{L\left[x_{LH}(\omega_2)\right]}
\end{eqnarray*}
 and obtain from~(\ref{eq:dispUH}),~(\ref{eq:dispLH}),~(\ref{eq:eq})
\begin{equation}
k_0^2(x_1)=k_2^2(x_1)=\frac{\tilde\varkappa_*^4}{\varkappa_{UH}^2},\qquad
k_0^2(x_2)=k_2^2(x_2)=\varkappa_{UH}^2\label{eq:decpoints}
\end{equation}
where following parameters are introduced
\begin{eqnarray*}
\fl\tilde\varkappa_*^4=\frac{\lambda_2^3\varkappa_*^4+\lambda_0^3k_*^4}{\lambda_2^3-\lambda_0^3}\simeq
\varkappa_*^4,\qquad
\lambda_0^3=L\left[x_{UH}(\omega_0)\right]\ell_T^2(\omega_0),\qquad
\lambda_2^3=L\left[x_{LH}(\omega_2)\right]\ell_T^2(\omega_2)
\end{eqnarray*}
and $\varkappa_{UH}$ denotes the largest solution of the equation
\[
\varkappa_{UH}^2+\frac{\tilde\varkappa_*^4}{\varkappa_{UH}^2}=
\frac{x_{UH}(\omega_0)-x_{LH}(\omega_2)}{\lambda_2^3-\lambda_0^3}
\]

Equations~(\ref{eq:decpoints}) determine in the indistinct form
decay point positions in question. They allow us to calculate the
parameters necessary for formulation of threshold power equation.
In particular, the phase, gained in the feedback loop, can be
represented as
\[
\Phi=\frac23\lambda_2^3\left(k_2-k_1\right)^3+\frac\pi2
\simeq\frac23\lambda_2^3\frac{\left(\varkappa_{UH}^2-\varkappa_*^2\right)^3}{\varkappa_{UH}^3}+\frac\pi2
\]

The length of the coherence region can be determined  as
\begin{equation}
\ell^2(x_{1,2})\approx\frac{2L(x_{1,2})\omega_0v_1\left(x_{1,2}\right)}{\omega_1^2-\omega_{ce}^2}
\label{eq:ell}
\end{equation}

Last important parameter is the value of LH wave group velocity in
the decay points. It can be estimated as
\[
-v_2(x_1)\simeq
v_2(x_2)\simeq\ell_T^2(\omega_2)\frac{2\omega_2\omega_{ce}^2}{\lambda_2\omega_{pe}^2}
\left[\frac32\left(\Phi-\frac\pi2\right)\right]^{1/3}
\]

We will be interested in the absolute instability threshold for
 mode $\Phi=2\pi$. This mode  has apparently almost the same threshold as fundamental mode $N=0$,
 which has the lowest one,
 but still can be described in WKB approximation. In
this case
\[
|Z(x_1)|\simeq
|Z(x_2)|\simeq-\ell(x_1)\ell(x_2)\nu_2(x_1)\nu_2(x_1)
\]
and to estimate the threshold we should solve an equation
\[
\frac{2\pi\rme^{\pi|Z|}}{|Z|\left|\Gamma(\rmi |Z|)\right|^2}=1
\]
which gives $|Z|\simeq0.110$. Substituting obtained expressions
for decay points~(\ref{eq:decpoints}) and coherence region
length~(\ref{eq:ell}) to~\eref{eq:nu} one obtains, that $|Z|$
should be calculated as
\[
|Z|=\left(\frac{e}{4m_e}\right)^2\frac{L^{4/3}\varkappa_*}{
\omega_{ce}^4\omega_0\ell_T^2(\omega_0)\ell_T^{4/3}(\omega_2)}\left[\frac32\left(\Phi-\frac\pi2\right)\right]^{-1/3}\frac{8P_i}{\rho^2}
\]
To obtain the threshold we have taken  here $|a_0|^2=8\pi
P(\omega_0^2-\omega_{ce}^2)/(\omega_0^2)$, where $P$ is the pump
wave power per unit square (in
$\mbox{erg}/(\mbox{s}\cdot\mbox{cm}^2)$).

Taking into account that for typical ST parameters in UHR
$\omega_0\sim\omega_{pe}$, and considering maximum
$k_z\sim\omega_2/(5V_{Te})$, when we can neglect LH wave Landau
damping, we obtain for $\Phi=2\pi$ a scaling for the threshold power
$P^*$
\[
P^*\left[\mbox{W}/\mbox{cm}^2\right]=1.4\cdot10^{-2}\left[\frac{\mbox{W}}{
\mbox{cm}^{2/3}\mbox{GHz}^{1/3}
\mbox{eV}^{11/6}\mbox{T}^{1/3}}\right]\cdot
\frac{f_0^{1/3}T_e^{13/6}B^{1/3}}{L^{4/3}}
\]
where $f_0[\mbox{GHz}]$ is the probing frequency, $T_e[\mbox{eV}]$
is the electron temperature, $L[\mbox{cm}]$ is the density
inhomogeneity scale~(\ref{eq:L}), $B[\mbox{T}]$ is the magnetic
field in plasma.

We calculate $P^*$ for MAST tokamak parameters:
$f_0=\omega_0/(2\pi)=57.5$~GHz, $T_e=100$~eV, $B=3.2$~kGs (in UHR
position), $L=5$~cm. In this case one obtains
$P^*=0.9\mbox{~MW}/\mbox{m}^2$, which gives for gaussian antenna
beam with radius $\rho=10$~cm threshold power $P_i^*\simeq30$~kW.
\section{Discussion}
\label{sec:disc} At first we discuss the approximations used. Our
analysis is performed in WKB approximation, which holds true, when
two following conditions are satisfied:
\begin{itemize}
\item Decay points $x_{1,2}$ are situated far enough from LH wave
turning point $x_*$. More accurately, taking into account that
electric field of LH wave in the vicinity of the turning point can
be expressed in terms of Airy function, it can be written as
\begin{equation}
x_{1,2}-x_*\gg\ell_A\label{eq:crit:WKB1}
\end{equation}
where Airy scale
$\ell_A=2^{2/3}\lambda_2$. In our case
\[
\frac{x_{1,2}-x_*}{\ell_A}\simeq\left[\frac34\left(\Phi-\frac\pi2\right)\right]^{2/3}
\]
and the condition~(\ref{eq:crit:WKB1}) can be  shown to be
satisfied even for $\Phi=2\pi$. \item Length of decay region is
not larger than Airy scale, which provides that all decay region
is situated far enough from the turning point. The coherence
region size~(\ref{eq:ell}) can be estimated as
$\ell\simeq\left(2\lambda_0^3\varkappa_*\right)^{1/2}$ and it can
be shown that condition
\begin{equation}
\frac{\ell}{\ell_A}<1 \label{eq:crit:WKB2}
\end{equation}
can be satisfied for wide range ST experiment parameters. \item
The distance between extraordinary wave cut-off and UHR, which can
be estimated as
\[
\Delta x\simeq\frac{\omega_{ce}}{\omega_0}L
\]
should be much larger than pump wavelength in the decay region.
 The last can
be estimated as $\Lambda_0(x_{1,2})\simeq2\pi/\varkappa_*$.
Corresponding condition, which provides WKB-representation of the
UH waves to be correct, takes the form
\[
\mu\equiv\frac{\omega_{ce}(x_{UH})L\varkappa_*}{2\pi\omega_0}\gg1
\]
This criterion is rather strict due to low magnetic field, which
is typical for ST. But it can be shown to be satisfied for MAST
experiment parameters, where $\mu\sim6$.
\end{itemize}
\begin{figure}
\begin{center}
\includegraphics[height=0.2\textheight]{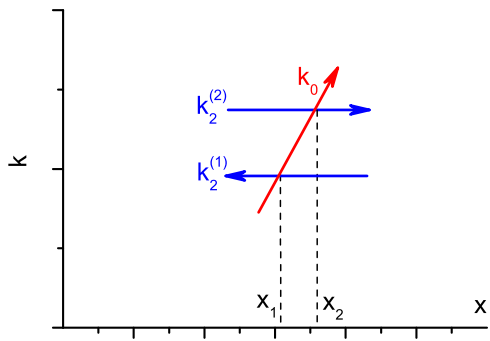}
\end{center}
\caption{\label{fig:disp2} Approximate representation of the
dispersion curves in the vicinity of decay points.}
\end{figure}

Our consideration, which is based on the formulae~(\ref{eq:S}),
seems to be sensitive to the possibility to consider decay points
as separate amplifiers of the incident wave. The condition for
that is $x_2-x_1\gg\ell$, which is equivalent to
\begin{equation}
\frac{x_2-x_1}{\ell}\simeq\left[\frac32\left(\Phi-\frac\pi2\right)\right]^{1/3}\frac{\ell}{\lambda_2}\gg1\label{eq:crit:3}
\end{equation}
Comparing that with~\eref{eq:crit:WKB2}, one obtains
that~\eref{eq:crit:3} can be satisfied for $\Phi=2\pi$ in rather
narrow range of parameters.  But, actually, an accurate analysis
shows, that for the dispersion curves behavior, which in the
region in question can be approximated as in~\fref{fig:disp2}, the
decay points joint influence is the same as given by our
consideration.
\section{Conclusion}
In the paper absolute instability of parametrical reflection of
upper hybrid wave  is analyzed in WKB approximation. The reflection
is assumed to be accompanied by radiation of lower hybrid wave.
Equations, describing the decay, obtained in $k\rho$-approximation.
The decay threshold is determined. It is shown that threshold power
increases with pump frequency and electron temperature. Threshold
power is estimated for typical parameters of experiment in MAST
tokamak. It is shown that in this case parametrical reflection
arises, if probing power exceeds $90\mbox{~W}/\mbox{cm}^2$, which
gives $30$~kW in a beam of $10$~cm radius.
\ack The support of RFBR grants 04-02-16404, 02-02-81033
(Bel~2002-a) is acknowledged. A.V.~Surkov~is thankful to
the``Dynasty'' foundation for supporting his research.
\end{document}